\documentclass[prl,nofootinbib,twocolumn,superscriptaddress]{revtex4}
\def\mysection#1{{\bf #1.} }

\usepackage{verbatim}
\usepackage{epsfig}
\usepackage{graphicx}

\newcommand{\Mp}{M_{\rm Pl}}

\begin{document}
{\hspace*{13cm}\vbox{\hbox{WIS/24/05-Sept-DPP}
    \hbox{hep-ph/0510186}}}

\vspace*{-10mm}

\title{\boldmath High Density Preheating Effects on Q-ball Decays and MSSM Inflation}

\author{Micha Berkooz}\email{micha.berkooz@weizmann.ac.il}
\affiliation{Department of Particle Physics,
  Weizmann Institute of Science, Rehovot 76100,
  Israel}

\author{Daniel J.H. Chung}\email{danielchung@wisc.edu}
\affiliation{Department of Physics, University of Wisconsin,
Madison, WI 53706}

\author{Tomer Volansky}\email{tomer.volansky@weizmann.ac.il}
\affiliation{Department of Particle Physics,
  Weizmann Institute of Science, Rehovot 76100, Israel}

\vspace*{1cm}

%







\begin{abstract}
Non-perturbative preheating decay of post-inflationary condensates
often results in a high density, low momenta, non-thermal gas. In the
case where the non-perturbative classical evolution also leads to
Q-balls, this effect shields them from instant dissociation, and may
radically change the thermal history of the universe. For example, in
a large class of inflationary scenarios, motivated by the MSSM and its
embedding in string theory, the reheat temperature changes by a
multiplicative factor of $10^{12}$.
%
%
%
\end{abstract}



\maketitle

%


\mysection{Introduction}
\label{sec:introduction}
One of the dominant paradigms of modern cosmology is that of
inflation. In recent years much emphasis has been given to the epoch
immediately after inflation, in an effort to connect it to the
Standard Model (SM) or its possible extensions, such as the MSSM.
Although this goal has not been achieved yet, much progress has been
made in understanding the relevant processes of reheating and
preheating.

While instabilities of the condensate may lead to intensive
production of particles\cite{Kofman:1997yn,Felder:2000hj}, it often also
generates non-topological solitons known as
Q-balls\cite{Lee:1991ax,Coleman:1985ki}. Such solitons are prevalent
in most extensions of the SM, and in particular in supersymmetric
ones.
For example, as noted by \cite{Enqvist:2000gq,Kusenko:1997zq}, the
existence of Q-ball solutions is generic in the MSSM parameter
space. Furthermore, due to inflation, we expect that some of the
condensates of these scalar fields will be displaced far from the
global minimum of their potential, thus generating the right initial
conditions for the formation of Q-balls.  The relevance of the
latter to cosmology depends on how long these Q-balls live.


Much like the approach to thermalization in preheating, Q-balls are
generated by dynamical instabilities acting on the condensate. In
fact, the point of view that we advocate in the current paper is
that Q-ball formation should be thought of as a more controllable
example of preheating in that the Lyapunov exponents around the
coherent condensate configuration could be smaller. A related fact
is that part of the systems falls into a new stable configuration
which is close by in phase space, namely Q-balls. As we point out,
yet another similarity is revealed when not all the energy of the
condensate ends up in the Q-balls. This is the generic case which is
most relevant to cosmology. In this case scaling arguments,
supported by simulations of Q-ball formation \cite{Kasuya:2000wx},
indicate that the remaining condensate go into a cold gas of low
momenta particles with large occupation number\cite{Berkooz:2005sf}.

%
%

This cold gas can decay into hot plasma, which in turn can bombard
and destroy the Q-balls. In the current paper we discuss the details
of this process and how the special off-equilibrium features of the
gas, such as its large number density, can actually stabilize
Q-balls at a large range of charges (for a review of the standard
computations of the Q-ball formation/destruction see
\cite{Enqvist:2003gh}).

The basic observation is that the finite density induces large
masses to all fields to which it couples, which are much larger than
termal masses for the same energy density. Thus, the gas's life-time
is significantly increased, thereby avoiding dissociation of the
Q-balls. Furthermore, this finite density-induced large mass can
also suppress the direct decay rate of Q-balls.  The upshot of all
this is that the longevity of Q-balls is generically increased by
the finite density effects. It should be stressed that finite
density effects, although similar in character, are not the same as
the effects of non-vanishing vacuum expectation value.

Relying on this observation and its consequence of a low reheat
temperature, one can for example rule out flat directions in modular
inflationary scenarios with gravity mediated SUSY breaking, in which
the flat directions have a displacement of order $\Mp$ from the
minimum.  The use of such moduli is very common, and is almost
inevitable \cite{Berkooz:2004yc}, for example, in models of hybrid
inflation with gravity mediated SUSY breaking. As detailed in
\cite{Berkooz:2005sf}, this leads to a unique MSSM flat direction,
$H_u H_d$, which is not carrying $B-L$ charge and can be involved in
modular inflation. This information is extremely useful for building
any model of inflation that involves the MSSM fields or those of
other supersymmetric extensions.

The purpose of current letter is to consider the effects of the
large occupation number of the gas on Q-balls. In addition to the
case in \cite{Berkooz:2005sf}, in which Q-giants with charge $Q\sim
10^{28}$ were considered, we show that the longevity of the Q-balls
is modified all the way down to charges of $10^{14}$, where the
effect becomes of order 1.



\mysection{High density effects and MSSM inflation}
\label{sec:Highdensity}
The scenario that we will focus on is the case of thick wall Q-balls
forming from a fragmenting condensate of a scalar field $\phi$. For
example, this will be the case in gravity mediated SUSY breaking with
the $U(1)$ invariant potential
\begin{equation}
\label{eq:ptntl}
U(\phi)=m_\phi^2 \left[ 1+ K \ln
\left(\frac{|\phi|^2}{M^2} \right) \right] |\phi|^2.
\end{equation}
Q-balls form for $K<0$, and the Q-ball solution is given
by\cite{Enqvist:1998en}
\begin{equation}
\phi(r)=\phi_0 \exp[-r^2/R^2] \exp[i w t],
\end{equation}
where its radius, frequency, mass, and $U(1)$ charge are given by
\begin{eqnarray}
R& =& \sqrt{\frac{2}{|K|}} m_\phi^{-1}\\
w^2 &=& m_\phi^2 \left[ 1- 2K + K  \ln \left(\frac{\phi_0^2}{M^2}
  \right) \right]\\
M_Q &=
&\frac{2\pi^{3/2}}{|K|^{3/2}} \frac{\phi_{0}^{2}}{m_{\phi}}\left[ 1-K(\frac{5}{2}-\ln[\frac{\phi_{0}^{2}}{M^{2}}])\right]\\
Q& =& \frac{2 \pi^{3/2}}{|K|^{3/2}}
\frac{\phi_0^2}{m_\phi^2}\left[1-K+\frac{K}{2} \ln
\left(\frac{\phi_0^2}{M^2}
  \right)\right].
\end{eqnarray}
Hence, when a condensate with a large VEV, $ \phi \gg m_\phi$, breaks
up into Q-balls with a core VEV of order $\phi_0\lesssim \phi$,
Q-balls have a large charge.

In the case that most of the energy in the universe is stored in the
fragmenting condensate, as in the post-inflationary scenario of
\cite{Berkooz:2005sf}, a significant fraction of the charge and
energy ends up inside the Q-balls\cite{Kasuya:2000wx}. The energy per
unit charge is determined by the eccentricity of the orbit. It is
$m_\phi$ if the eccentricity is zero, and larger otherwise which is the
generic case. If the eccentricity is smaller than one, but of order
one, all the charge falls into Q-balls (which can subsequently be
eroded by collisions), and the remaining energy will form a cold,
high density gas.  The number density in the gas, denoted by
$\delta\phi$, is approximately $n_1\sim m_\phi \langle
|\delta\phi|^2 \rangle$. Note that this density can be quite large
since $\langle |\delta\phi^2|\rangle\sim \phi_0^2$ despite the fact
the $U(1)$ charge outside of the Q-balls is negligible compared to
the charge inside.

This scenario is verified by simulations \cite{Kasuya:2000wx}, but
can be understood by the following scaling argument. Due to the
attractive forces in the condensate, the inhomogeneities grow
like\cite{Enqvist:2003gh} $\exp(\sqrt{|K|k^2}t)$, for
$|k|<\sqrt{2|K|}m_\phi$. Q-balls form when the shortest wave length
become non-linear. At this point, this is the smallest wave-length
scale in the non-linear problem and hence it also determines the
shortest wavelength of the gas which results from the same
non-linear dynamics.

This result seems to be ubiquitous to non-perturbative preheating
processes \cite{Kofman:1997yn,Felder:2000hj}. Both in parametric
resonance and in tachyonic preheating, the non-linear processes are
characterized by a low UV cut-off on the modes that are excited. In
parametric resonance, the cut-off originates from the frequency of
oscillation of the driving field, and in the tachyonic preheating it
is set by the negative mass-squared of the tachyon. In the next
section we will discuss the approach to equilibrium in our model.

Had there been hot plasma between the Q-balls, it would have
bombarded the Q-balls, leading to their rapid decay by dissociation.
For a thermal gas between the Q-balls this would have been the case.
For example a thermal gas originating at $\phi_0 \sim 0.1 \Mp$ (and
$m_\phi\sim 10^3$ GeV) would have a temperature of $10^9$ GeV which
would dissociate the Q-balls rapidly.

This is where the high density effects become important, as they
prevent the cold gas outside the Q-balls from decaying instantly due
to its large density. Indeed, this large density contributes to the
masses of all particles that $\phi$ couples to. Collectively
denoting these particles by $X$, their effective mass is given by,
\begin{equation}
m_X \sim \sqrt{\langle |\delta\phi|^2 \rangle} \sim \phi_0 \propto
m_\phi \sqrt{Q}.
\end{equation}
Note that this induced mass is due to the large variation of the
field, despite the fact that $\langle\delta\phi\rangle =0$. Thus the decay
of the gas is mediated by fields of large mass and can therefore be
estimated by
\begin{equation}
  \label{eq:2}
\Gamma_{\rm gas} \simeq \frac{m_\phi^3}{8\pi \langle
|\delta\phi|^2\rangle},
\end{equation}
(again although $\langle\delta\phi\rangle=0$). This leads to a much
colder plasma outside the Q-balls than the naive estimate. Indeed,
one finds using (\ref{eq:2}), that for Q-balls with charge
\begin{equation}
  \label{eq:5.1}
  Q \gtrsim Q_{\rm min} \equiv 5\times 10^{12}
  \left(\frac{m_\phi}{10^3
  \mbox{GeV}}\right)^{-2/3} \left(\frac{|K|}{0.01}\right)^{-3/2},
\end{equation}
the intermediate temperature of the plasma - after the gas decays
and before the Q-balls' final decay and subsequent reheat of the
universe - is given by
\begin{equation}
  \label{eq:4}
  T_{\rm inter}\simeq 10^5 \left(\frac{m_\phi}{10^3
      \mbox{GeV}}\right)^{5/6}  \left(\frac{g_*}{200}\right)^{-1/4} \mbox{GeV}.
\end{equation}
Smaller Q-balls are not generated since the decay rates of both the
gas and of the Q-balls themselves are smaller than the expansion
rate, rendering such objects unstable.

On the other hand, the temperature above which dissociation takes
place was found to be \cite{Enqvist:1998en,Berkooz:2005sf},
\begin{equation}
T_{\rm diss} \simeq 10^5  \left(\frac{|K|}{10^{-2}}\right)^{1/2}
\left(\frac{200}{g_*}\right)^{1/4} \left(\frac{m_\phi}{10^3
    \mbox{GeV}}\right) \left( \frac{Q}{10^{14}}\right)^{1/4}\mbox{GeV}.
\end{equation}
Hence, contrary to naive expectations, for Q-balls with charge
$\gtrsim 10^{14}$, amounting to the majority of stable Q-balls in the
spectrum, the decay of the gas cannot cause dissociation of these
objects.  We stress, once more, that this result is due to the high
density of the non-thermal gas.  We also note that in the results
presented here we have neglected collisions of Q-balls, which do not
change the above conclusions. For the case of Q-giants this was shown
in \cite{Berkooz:2005sf}, but it is true more generally.

In particular, Q-giants, with charge of order
$10^{28}$ are not dissociated and survive well into nucleosynthesis.
As shown below, the reheat temperature after the final decay
of the Q-giants is given by\cite{Berkooz:2005sf}
\begin{eqnarray}
 \label{eq:3}
 \begin{array}{ll}
 T_{\rm RH} =
    10^{-3}&{\displaystyle \left(\frac{m_\phi}{10^3 \rm{GeV}}
  \right)^{1/2}\left(\frac{Q}{10^{28}} \right)^{-1/2}\times
  }
  \\
  &{\displaystyle \left(\frac{|K|}{0.01}\right)^{-3/4}\left(\frac{g_*}{200}\right)^{-1/4} \mbox{GeV}. }
  \end{array}
\end{eqnarray}
In this case, the high density effects results in a reheat
temperature twelve orders of magnitude smaller!

As promised in the introduction, in the case of Q-giants generated
along the flat directions of the MSSM, such considerations exclude
all flat directions with $K<0$ from having displacement of order
$M_p$. One can further exclude $K>0$ flat directions with lepton or
baryon number. This leads to a unique viable flat direction in the
MSSM, $H_u H_d$, pinpointing the inflationary sector in the MSSM.
%

\mysection{The approach to thermal equilibrium} Once formed, the
non-thermal gas slowly approaches thermal equilibrium.  In some
respects, the relevant processes are similar to those discussed in
thermalization of the usual preheating scenario\cite{Felder:2000hr},
however, there are some crucial differences. One such difference is
that during the formation of Q-balls, no large number densities of
fields other than $\phi$ are present. In \cite{Felder:2000hr} such
fields are initially populated via rapid parametric resonance or
tachyonic preheating. In our case, the $\phi$ VEV gives mass to a
heavy particle $X$ (which in turn may couple to some particles). For
generic trajectories of the complex $\phi$ VEV, with eccentricity of
order one, the mass of $X$ is much larger than the period of
oscillation of $\phi$ (of order $m_\phi$) and hence cannot be
excited by parametric resonance. Also, in (\ref{eq:ptntl}) there is
no field that can undergo tachyonic preheating.

%
%

At large occupation number, the number density of a thermalized gas
behaves as $n_k \propto 1/k$ for $k \gg m$ and $n_k = const$ for $k
\ll m$.  In our case, initially the gas is highly populated at
momenta $k_* \sim \sqrt{K}m_\phi \ll m_\phi$ and so it's spectrum is
sharply cut off at $k_*$.  During thermalization two distinct
processes occur: (i) At low momenta, the number density evolve to
become constant as in the thermal distribution.  (ii)  Higher
momentum modes become populated. As in
\cite{Kofman:1997yn,Felder:2000hr}, the rate of the former is
enhanced due to the large occupation number at low momenta.  Thus
the relaxation time for low momentum is small.  The production of
high momentum modes, however, must involve higher order mode-mode
interactions (i.e. strong turbulence), which are not enhanced since
the high momentum modes have low occupation number.

Relying on the these observations, we can integrate out $X$ and
obtain that the relevant decay rate for these many-body processes is
suppressed by the mediation of heavy particles and can be estimated
by,
\begin{equation}
  \label{eq:1}
  \Gamma_{\rm thermal} = n_\phi\langle\sigma_\phi v\rangle \lesssim
  \frac{m_\phi^3}{8\pi\langle|\delta\phi|^2\rangle}.
\end{equation}
This estimate is not limited to a $2\rightarrow 2$ (which actually
hardly contributes to populating high momentum modes), but is true
for all $n$-body interactions for the specific potential
(\ref{eq:ptntl}). Hence we expect it to be valid in the large
occupation number case at hand.

Eq. (\ref{eq:1}) is of the order the decay rate of the gas,
eq.~(\ref{eq:2}). Hence, while thermalization of low momentum modes
occurs quickly, the gas never reaches true thermal equilibrium and
decays before it has time to populate high momentum modes.  Had
thermalization occurred faster, considerably higher temperature and
consequent dissociation of the Q-balls would have resulted.

\mysection{Q-ball decay} The computation of the decay rate of the
Q-balls also require modification in the presence of the non-thermal
gas (although the effect is much less pronounced than the
elimination of the high temperature dissociation process as
discussed above). It also clarifies some of the physics involved.
Hence below we reproduce the corrected decay rate and reheat
temperature of the Q-balls, taking into account the effects of the
gas.  We will mostly follow the techniques of
Ref.~\cite{Enqvist:1998en}.

In the core of the Q-ball the VEV of $\phi$ is large, the masses of
$X$ are large and the decay is suppressed.  This is the familiar
computation \cite{Cohen:1986ct,Enqvist:1998en,Enqvist:2002rj}.
However, outside of the core, the novel feature is that due to the
high density of the gas, $X$ is massive even in the region between
the Q-balls where the coherent VEV $\langle\phi\rangle$ vanishes,
again suppressing the decay through the surface of the Q-ball.  In
the usual computation, one defines the core radius $r_{c}$
\begin{equation} r_{c}=R\sqrt{-\ln[m_{\phi}/(g\phi_{0})]}\equiv
\gamma_c R.
\end{equation}
such that the decay rates are
\begin{eqnarray}
&\Gamma(r<r_{c})&\sim m_{\phi}^{3}/ 8\pi |\phi(r)|^2,\\
&\Gamma(r>r_{c})&\sim g^2 m_\phi/ 8\pi.
\end{eqnarray}
In our case, there is a different time dependent scale $r_d$ which
replaces $r_c$. The reason for this is that in a mixed Q-ball/cold gas
system the mass of $X$ is $m_X^2 \sim g^2(|\langle\phi\rangle|^2 +
\langle|\delta\phi|^2\rangle)$. The latter contribution is given by $\phi_0^2 f(t)
(a_i/a)^3$, where $f(t)$ takes into account the decay of the gas
into light particles and mode-mode interactions, both of which reduce
$\langle|\delta \phi|^2\rangle_{\rm gas}$. Following the discussion above, it
may be approximated by 1 until the gas decays after which it rapidly
approaches 0. We define $r_d$ to be the location where these two
contributions are equal
\begin{equation}
r_d= R\, \mbox{min}\left(\sqrt{ {1\over 2}  \ln{( (a/a_i)^3 f/g^2)}},
\gamma_c\right) \equiv \gamma_d R.
\end{equation}
The decay rates are now given by
\begin{eqnarray}
&\Gamma(r<r_d(t))&\sim m_{\phi}^{3}/ 8\pi |\phi(r)|^2,\\
&\Gamma(r>r_d(t))&\sim g^2 m^3_\phi/ 8\pi (m_\phi^2 + \phi_0^2 f(t)
(a_i/a)^3).
\end{eqnarray}
The above has the correct limit, $f\rightarrow 0$, in the case the gas
quickly decays and one returns to the usual vacuum decay rate of
the Q-balls.
%

The reheating temperature at the end of Q-ball decay can be
evaluated as follows.
The effective decay rate of the Q-ball is obtained from the charge depletion rate,
\begin{eqnarray}
\label{eq:partonicdesc} && \Gamma_Q= {1\over Q}\left| \frac{dQ}{dt}
\right| = {1\over Q} \int
2w\phi^{2}(r)\Gamma(r)4\pi r^{2}dr \\
&& \simeq\frac{w
  m_{\phi}^{3}R^{3}}{Q}
\left[ \frac{\gamma_d^3}{3} + \cal{O}(\gamma)
 \right]\nonumber
\end{eqnarray}
where we have only written the leading order term, given that
$\gamma_d>1$.  This leading contribution comes from the region
$r<r_d$. To this end the suppression of the decay rate relative to
the usual scenario occurs because $\gamma_d<\gamma_c$.  Since
however, $\gamma_d$ is logarithmic in time, this suppression is only
mild.  One way to see this is by solving the resulting Boltzmann
equations~\cite{Berkooz:2005sf}:
\begin{eqnarray}
\frac{d\rho_\gamma}{dt} +4 H \rho_\gamma & = & \int dQ f_Q(t,Q)
\Gamma_Q(t,Q)
M(Q) \\
\partial_t f_Q +3 H f_Q & = & Q \Gamma_Q(t,Q) \partial_Q f_Q(t,Q)
\end{eqnarray}
where $\rho_\gamma$ is the radiation energy density from the Q-ball
decay, $M(Q)$ is the mass of the Q-ball with charge $Q$, and
$f_Q(t,Q)$ is the Q-ball charge distribution function.  This was
done in \cite{Berkooz:2005sf}, where the solution to these equations
was presented in for case of Q-giants, without, however, accounting
for the new decay rates.  One may consistently treat $\gamma_d$ as
constant, and extract the time of decay by comparing $\Gamma_Q$ to
the Hubble scale $H$.  One finds
\begin{equation}
  \label{eq:5.2}
  \frac{t_{\rm decay}}{t_i} \sim \ \frac{\beta}{(\ln\beta)^{3/2}}, \;\;\;\; \beta
  \equiv \frac{\phi_0^3}{m_\phi^2\Mp}
\end{equation}
where $t_i$ is the time at which the Q-balls form.  This gives only a
suppression to $\Gamma_Q$ relative to the conventional computation, of
order $\gamma_d^3/\gamma_c^3 \simeq 1/5$ at the time of decay.
Plugging the above $\gamma_d$ in (\ref{eq:partonicdesc}) one obtains
the reheat temperature, eq.  (\ref{eq:3}).


\mysection{Conclusions}
\label{sec:conclusions}
In recent years, a theory of preheating has
emerged in an effort to understand universal features in the
transition between the phase of a coherent condensate right after
inflation, and the phase of thermal SM particles. In this letter we
explored aspects of this transition in a case where Q-balls dominate
an intermediate stage between the two phases.

We outlined key features of the non-thermal, high occupation number
gas that remain between Q-ball during their non-linear formation
process. We explained how such effects can change significantly the
Q-ball decay dynamics. Compared to the situation with a thermal gas
between the Q-balls, this high density leads to a suppression of up
to $10^{12}$ in the reheat temperature.

This, for example, occurs in inflationary models where inflation takes
place at large VEVs of scalars present in SUSY extensions of the SM,
such as the MSSM. In such cases, the Q-balls can attain charges of
order $10^{28}$. The non-thermal aspects of the gas suppress
dissociation processes, leading to a reheat temperature below the
nucleosynthesis temperature. Under realistic circumstances, this picks
out a unique MSSM flat direction $H_u H_d$ to be part of the inflaton
dynamics.

These finite density effects change the reheat temperature for much
smaller Q-balls as well, all the way down to a charge of
$10^{14}$.

\mysection{Acknowledgement}

The work of DJHC is supported by DOE Outstanding Junior Investigator
Award DE-FG02-95ER40896 and NSF grant PHY-0506002.  The work of TV
is supported by the Chlore foundation and by the United
States-Israel BSF. The work of MB is supported by the Israel Science
Foundation, by the Braun-Roger-Siegl foundation, by
EU-HPRN-CT-2000-00122, by GIF, by Minerva, by the Einstein Center
and by the Blumenstein foundation.

\end{document}